\documentclass[10pt,twocolumn,superscriptaddress,showpacs,floatfix]{revtex4}
\usepackage[dvips]{graphicx}
\usepackage{amssymb}
\newtheorem{theorem}{Theorem}[section]
\newenvironment{proof}{Proof}{$\hfill\Box$}
\begin{document}
\title{Condition numbers and scale free graphs}
\author{Gabriel Acosta}%
\email{gacosta@ungs.edu.ar}
\affiliation{Instituto de Ciencias, Univ. Nac. de Gral. Sarmiento}

\author{Mat\'\i as Gra\~na}%
\email{matiasg@dm.uba.ar}
\affiliation{Depto de Matem\'aticas, FCEyN, Universidad de Buenos Aires}

\author{Juan Pablo Pinasco}%
\email{jpinasco@ungs.edu.ar}
\affiliation{Instituto de Ciencias, Univ. Nac. de Gral. Sarmiento}

\date{Received: date / Revised version: date}
\begin{abstract}
	In this work we study the condition number of the least square matrix
	corresponding to scale free networks. We compute a theoretical lower bound of
	the condition number which proves that they are ill conditioned. Also, we
	analyze several matrices from networks generated with the linear preferential
	attachment model showing that it is very difficult to compute the power law
	exponent by the least square method due to the severe lost of accuracy expected
	from the corresponding condition numbers.
\end{abstract}
\maketitle
\section{Introduction}
\label{intro}

In the last years several networks were analyzed, like internet routers,
biological and metabolic networks, or sexual contacts \cite{FFF}, \cite{JTAOB},
\cite{LEASA}, and the node degree distributions of all of them seem to follow a
power law.  Also, several models of graph growth were presented in order to
explain the emergence of this power law distribution \cite{AB}, \cite{KR},
\cite{DMS}. However, several critics appeared, mainly focusing on sampling bias
\cite{ACKM},  \cite{LKJ}, \cite{PR}, \cite{CCG}, \cite{CM}, \cite{AAB},
\cite{LBCX}, and the quality of data fitting \cite{HJ}, \cite{KW}, \cite{SMA}.

Recently, a simple experiment was presented in \cite{GMY} studying the linear
fit on the log log scale of computationally generated data with a pure power
law distribution, and a severe bias error was reported ($36\%$, and $29\%$ with
logarithmic bins).

In this work we present an underlying problem which explains those errors:
regrettably, the matrix in the least square method is ill conditioned. Let $n$
be the maximum degree of the network, we show that the condition number grows
at least as the logarithm of $n$. Moreover, we introduce a parameter $c\in [0,1]$
and we consider only the node degree distribution on $[cn, n]$ (in fact, this
is a usual procedure, see \cite{N}). Numerical computations show that the
situation is worse when we focus on the tail of the distribution.

Our results complement the ones in \cite{KW}, where biological networks were
considered and a different statistical problem arose, since on that work the
power law fit was performed with the maximum likelihood method.

Also, we compute the matrix condition for scale free graphs generated with the
linear preferential attachment model introduced by Barabasi and Albert
\cite{AB}.  We show that the matrix condition grows when the network size
increases.

\section{Main Results}
\label{sec:1}
\subsection{Condition Number}
\label{subsec:1}

For a given matrix $A \in R^{m\times m}$, and a matrix norm $\|.\|$, the
condition
number is defined as
$$cond(A)= \|A\|\|A^{-1}\|, \qquad cond(A)=\infty \, \mbox{ if } \,
det(a)=0$$
Usually, for the $2$-norm the condition is denoted $cond(A)_2$. The $2$-norm
is an
operator type norm, i.e. for $v\in R^m$, taking the vectorial Euclidean norm
$$\|v\|_2:= \left( \sum_{i=1}^{m} |v_i|^2\right)^\frac12$$
we have
$$\|A\|_2=sup \{\|Av\|_2\ \ :\ \ \|v\|_2=1\} .  $$

Concerning the condition number, the following results are well known
\cite{GV}:
\begin{equation} \label{condi} cond(A)_2 =
\frac{\lambda_{max}}{\lambda_{min}}.
\end{equation}
where $\lambda_{min}$ and $\lambda_{max}$ are the minimum and the maximum eigenvalue
(in absolute value), and
\begin{equation}
\frac{1}{cond(A)_2}
    =inf \left\{ \frac{\|A-S\|_2}{\|A\|_2} \ :\ S \ \ \mbox{singular}
\right\} \label{dist}
\end{equation}
which says that $cond(A)_2 $ is the reciprocal of the relative distance of
$A$ to the
set of singular matrices.

The interest in the condition number for matrices is related to the accuracy of
computations, since it gives a bound for the propagation of the relative error
in the data when a linear system is solved. If $cond(A) \sim 10^k$, then $k$ is
roughly the number of significant figures we can expect to lose in
computations.

More precisely, for  a general system $Ax=b$, if we consider a perturbation on
the right hand side $\tilde b$, then calling $\tilde x$ to the exact solution
of $A\tilde x=\tilde b$ it can be shown that
$$
\frac{\|x-\tilde x\|_2}{\|x\|_2}
	\le cond(A)_2\frac{\|b-\tilde b\|_2}{\|b\|_2}.
$$

A practical rule in statistics is to avoid the least square method when the
condition number is greater than or equal to $900$ (indeed they define
$\kappa(A)=cond(A)^{1/2}$, and $\kappa\ge 15$ is a strong sign of collinearity,
see for example \cite{CHP}).

\subsection{Theoretical Results}
\label{subsec:2}

Let us consider a graph $G$ with $k$ nodes $x_1, \cdots, x_k$, and $d(x_i)$ is
the degree of node $x_i$, that is, the number of links emanating from $x_i$.
Let us define
$$n=max\{d(x_i) \, : \, 1\le i \le k\}.$$
For each $j$, $1\le j \le n$, let $h_j$ be the number of nodes with degree
$j$. The existence of a power law dependence $h(d) = ad^{\gamma}$ is
usually observed in a  log-log plot, and computed with the least square method
after a logarithmic change of variables.

First we assume that the degrees span the full integer interval $[1,n]$. In
this case the matrix $A_n$ corresponding to the least square fit, regardless of
the measured data, is given by
$$A_n = \left(
\begin{array}{cc}
  n & \sum_{j=1}^n \ln(j)  \\
  \sum_{j=1}^n \ln(j)  & \sum_{j=1}^n \ln^2(j)  \\
\end{array}
\right) $$ In certain a sense, this correspond to the best situation where the
data span the full range of variables. The following result estimates the
condition number of $A_n$, when $n\to \infty$:

\begin{theorem}\label{th:uno}
For $n$ large, it holds
$$cond(A_n)_2\sim \ln^4(n) $$
\end{theorem}

\begin{proof}:
We use here (\ref{condi}). A straightforward computation of the eigenvalues
of $A_n$
gives
\begin{equation}
\label{maximo} \lambda_{max}=\Big(n+\sum_{j=1}^n \ln^2(j)\Big)+
\sqrt{\Delta}
\end{equation}
\begin{equation}
\label{minimo} \lambda_{min}=\Big(n+\sum_{j=1}^n
\ln^2(j)\Big)-\sqrt{\Delta},
\end{equation}
where
$$\Delta =
\Big(n-\sum_{j=1}^n \ln^2(j)\Big)^2+4\Big(\sum_{j=1}^n \ln(j)\Big)^2.$$

 For $n$
large we can write
$$\sum_{j=1}^n \ln(j) \sim n(ln(n)-1))+ O(ln(n)) $$
and
$$\sum_{j=1}^n \ln^2(j) \sim n(ln^2(n)-2ln(n)+2)+ O(ln^2(n)).$$
Replacing this expressions in (\ref{maximo}) and (\ref{minimo}), we get by
taking
limit $$lim_{n\to \infty}
\frac{\frac{\lambda_{max}}{\lambda_{min}}}{ln^4(n)}=1$$

\end{proof}

\bigskip

Since in practice logarithmic bin is preferred (see for example \cite{N}),
due to the
sparsity of measurements at the tail of the distribution, our next result
shows that
also the corresponding matrix is ill conditioned. We suppose that the
selected degrees
for the computation are of the form $e^j$ with $1\le j \le n$. Calling
$A_{e^n}$ the
corresponding least square matrix, we can write

$$ A_{e^n} = \left(
\begin{array}{cc}
  n & \sum_{j=1}^n j  \\
  \sum_{j=1}^n j   & \sum_{j=1}^n j^2  \\
\end{array}
\right) = \left(
\begin{array}{cc}
  n & \frac{n(n+1)}{2} \\
  \frac{n(n+1)}{2}   &\frac{n(n+1)(2n+1)}{6}   \\
\end{array}
\right).$$

And the following holds

\begin{theorem}\label{th:dos} For $n$ large $$cond(A_{e^n})_2\sim
\frac{4}{3} n^2.$$
\end{theorem}

\begin{proof}:
Using again (\ref{condi}), and computing explicitly the eigenvalues of
$A_{e^n}$, we
have

$$ \frac{\lambda_{max}}{\lambda_{min}} = \frac{7+2n^2+3n+
\sqrt{61+25n^2+42n+4n^4+12n^3}}{7+2n^2+3n-\sqrt{61+25n^2+42n+4n^4+12n^3}}
$$

Hence, for $n$ large
$$ cond(A_{e^n})_2= \frac{\lambda_{max}}{\lambda_{min}} \sim
\frac{4}{3}n^2.$$
\end{proof}

Numerical experiments in the next section suggest that considering a
logarithmic bin of the form $a e^j$ is unnecessary, since the condition number
grows almost independently of $a$, see Table  \ref{tab:1}.

\subsection{Numerical Simulations}\label{subsec:3}

In this section we present several numerical computations of matrix
conditions.

\begin{figure}
		\includegraphics[scale=.5]{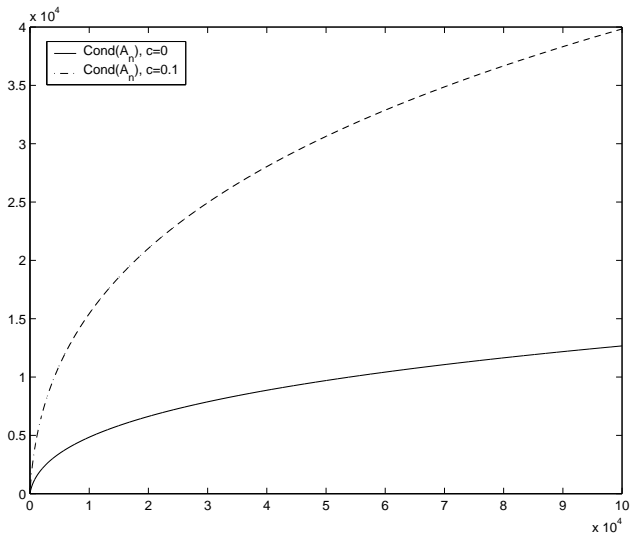}
	\caption{Condition number of $A_n$ with $n \le 10^5$} \label{fig:conda}
\end{figure}

We computed the condition number of matrix $A_n$  numerically by using MATLAB. Also,
we computed the condition number for the truncated matrix $A_{n}$, for each $n$ we
consider the matrix obtained with degree values between $cn$ and $n$. The results are
shown in Figure \ref{fig:conda} for $n \le 100000$, $c=0$ and $c=0.1$.

\begin{figure}
		\includegraphics[scale=.5]{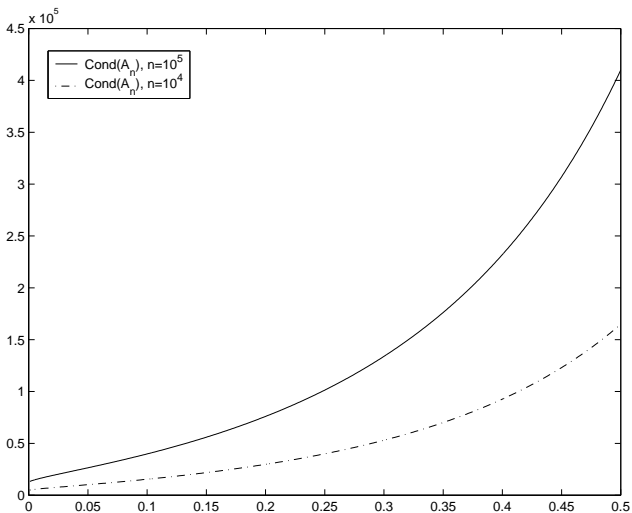}
	\caption{Condition number of $A_n$ with $0 \le c \le 0.5$}
	\label{fig:asubc3}
\end{figure}

We show the dependence on $c$ in Figure \ref{fig:asubc3}, for $n=10^4$ and
$n=10^5$,
with $c$ from $0$ to $0.5$.

In Table \ref{tab:1}  we show the condition numbers for logarithmic bins of
the form
$ae^j$, $1\le j \le n$, for $n=10^3, 10^4, 10^5,$ and $ 10^6$; and $a=0.1$,
$a=1$ and
$a=2$.

\begin{table}
\caption{Condition number with logarithmic bins}
\label{tab:1}
\begin{tabular}{cccc}
\hline\noalign{\smallskip}
$ae^j$, $1\le j \le n$ & a=1 & a=0.1 & a=2  \\
\noalign{\smallskip}\hline\noalign{\smallskip}
 $n=10^3$  & $1.319 \times 10^6$      & $1.337 \times 10^6$    & $1.343
\times 10^6$     \\
  $n=10^4$ &  $1.332 \times 10^8$     &  $1.334 \times 10^8$    &  $1.334
\times 10^8$     \\
 $n=10^5$ &  $1.333 \times 10^{10}$  &  $1.333 \times 10^{10}$ &  $1.333
\times 10^{10}$   \\
  $n=10^6$ &  $1.333 \times 10^{12}$  &  $1.333 \times 10^{12}$ &  $1.333
\times 10^{12}$   \\
\noalign{\smallskip}\hline
\end{tabular}
\end{table}

Finally, we consider the Linear Preferential Attachment model of Barabasi and
Albert.  This is a model of network growth, where a new node is added with a
link to a previously added node, chosen at random with a probability
proportional to its degree.

\begin{figure*}
		\includegraphics[scale=1]{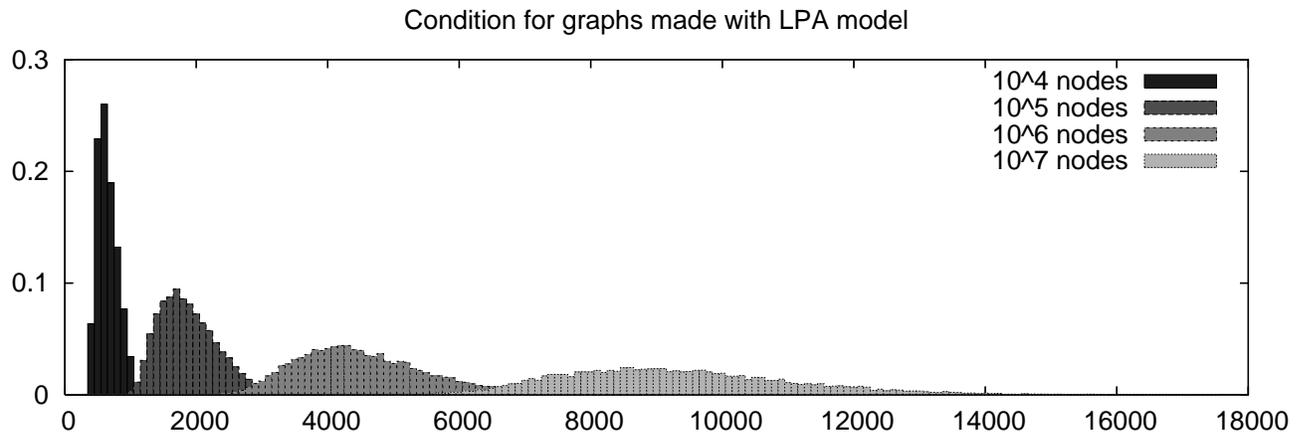}
	\caption{Condition number of graphs of $10^4$, $10^5$, $10^6$  and $10^7$
	nodes, computed over $5\times 10^4$, $2.5\times 10^4$, $10^4$ and $10^4$ graphs}
	\label{fig:ce}
\end{figure*}

We generated $5\times 10^4$ graphs of $10^4$ nodes, $25\times 10^3$ graphs of $10^5$
nodes, $10^4$ graphs of $10^6$ nodes, and $10^4$ graphs of $10^7$ nodes, and computed
the condition of the least square matrix associated with each one. We show the
distribution of values of the condition number in Figure \ref{fig:ce}. Also, in Table
\ref{tab:2}  we present the computation of mean values of the condition number for
$c=0$, $c=0.05$ and $c=0.1$.

\begin{table}
\caption{Mean value of condition numbers for LPA graphs with different values of $c$}
\label{tab:2}
\hspace{.6cm} \begin{tabular}{ccccc} \hline\noalign{\smallskip}
  Nodes &  Graphs & c=0 & c=0.05 & c=0.1 \\
  \noalign{\smallskip}\hline\noalign{\smallskip}
  $10^4$ & $5\times 10^4$   &  $113.7$ & $379.7 $ & $ 703.7$ \\
  $10^5$ & $2.5\times 10^4$ &  $223.5$ & $1058.4$ & $1928.8$ \\
  $10^6$ & $10^4$           &  $409.0$ & $2648.5$ & $4560.0$ \\
  $10^7$ & $10^4$           &  $703.8$ & $5897.6$ & $9369.5$ \\
\noalign{\smallskip}\hline
\end{tabular}
\end{table}

\section{Conclusions}

We have studied  the condition number of the least square matrix corresponding
to scale free networks. We computed theoretical lower bounds of the condition
numbers showing that it behaves roughly as the logarithm of the maximum degree
of the network, and numerical simulations support this fact. We also showed
that neglecting the less connected nodes of the network (a usual practice in
fact, since the interest is on the tail) things become even worse. Similar
conclusions can be drawn for the logarithmic bin.

Finally, for random networks generated with the Linear Preference Attachment
model, numerical computations of the condition numbers showed a severe ill
condition of the least square matrices, even for small sized networks ($10^4$
nodes). Clearly, in this context it is very difficult to compute the power law
exponent by the least square method due to the lost of accuracy expected from
the corresponding condition numbers.

\bigskip
\section*{Acknowledgements}

GA and JPP are partially supported by Fundacion Antorchas and ANPCyT. MG is
partially supported by Fundacion Antorchas and CONICET.

\end{document}